\documentclass[twocolumn,pra,showpacs,superscriptaddress]{revtex4-1}
\usepackage{amssymb}
\usepackage{amsmath}
\usepackage{graphicx}
\usepackage{subfigure}
\usepackage{natbib}
\usepackage{epsfig}
\usepackage{amsfonts}
\usepackage{mathrsfs}
\usepackage{CJK}
\usepackage{xcolor}
\usepackage[toc,page,title,titletoc,header]{appendix}

\begin{document}

\title{Sudden transition from naked atom decay to dressed atom decay}
\author{Wei Zhu}

\affiliation{Institute of Physics, Beijing National Laboratory for
  Condensed Matter Physics, Chinese Academy of Sciences,Beijing
  100190, China}

\author{D. L. Zhou}

\email{zhoudl72@iphy.ac.cn}

\affiliation{Institute of Physics, Beijing National Laboratory for
  Condensed Matter Physics, Chinese Academy of Sciences,Beijing
  100190, China}

\date{\today}

\begin{abstract}
  The studies on quantum open system play key roles not only in
  fundamental problems in quantum mechanics but also in quantum
  computing and information processes. Here we propose a scheme to use
  a one dimensional coupling cavity array (CCA) as an artificial
  electromagnetic environment of a two-level atom. For a finite length
  of CCA, we find that after a turning time the population of excited
  state deviates suddenly from the exponential decay. We show that
  physically this phenomena corresponds to a transition from a naked
  atom decay to a dressed state decay. We hope that our finding will
  promote the studies on quantum system with a finite size
  environment.
\end{abstract}


\maketitle


\section{Introduction}
\label{sec:introduction}

In nature there does not exist a completely closed quantum system. Any
quantum system must interact with environment, and the environment
in most cases is very complex and often not precisely specified, which
makes it extremely difficult to study such an open system. One obvious
reason for us to study an open system is that the knowledge on the
system's state can not be obtained without the interaction between the
measurement apparatus and the system. In addition, to control a
quantum system to implement some quantum computing and information
processes, we have to fight against the decoherence and dissipation
induced by the environment~\cite{book}.

Our aim is to study the effect of an environment with finite size. We
model an environment with a one-dimensional coupling cavity array,
which may be realized in experiments by coupled superconducting
transmission line resonators or defect resonators in photonic
crystals~\cite{e1,e2,e3,e4}. In literature, a one-dimensional coupling
cavity array with a two-level atom in is first introduced to study
single photon scattering problem~\cite{a1}. Then many variants of this
model are studied in recent years, for example, a super cavity can be
formed by embedding two atoms to control the tunneling or directly
decreasing two tunneling strengths~\cite{b1,b2}. In addition, when
putting an excited atom in the middle cavity of the CCA, the atom
finally will evolve into two bound states and illustrate the
oscillation behavior~\cite{a2,a3}.

Here we study the spontaneous decay of a two-level atom located in one
node of a one dimensional (1D) coupling cavity array (CCA). For the
atom, the 1D CCA can be regarded as a 1D electromagnetic environment.
Because the length of the cavity array can be controlled in experiments,
we have the opportunity to study the effect of an environment with
finite size. For an environment with finite size, we may expect the
spontaneous decay of the atom will not be exponential one~\cite{a4,a5,a6,a7,a8}. We find
that there is a sudden deviation of the exponential decay at some
turning time. We theoretically explain this phenomena as the
transition from a naked atom decay to a dressed atom decay, and
numerical results confirm this underlying mechanism.

The remainder of the paper is organized as follows. In Sec. 2, the
theoretical model of a two-level atom in a 1D CCA is introduced as a
microscopic model of an open quantum system. In Sec. 3, we analyze the
condition for the exponential decay in our system, and gives the
analytical result on the decay rate. In Sec. 4, we give the numerical
results on the atom decay for the CCA with finite length, and find
that the deviation of the exponential decay is a sudden transition at
some turning time. We give a physical explanation why this happens and
predict the turning time. In Sec. 5, after the turning time, we gives
the results based on the dressed atom decay, which is confirmed by the
numerical results. Finally we will give some discussions and a brief
summary.

\section{Model: a two-level atom in 1D CCA}

\begin{figure}[!htbp]
  \includegraphics[width=8cm]{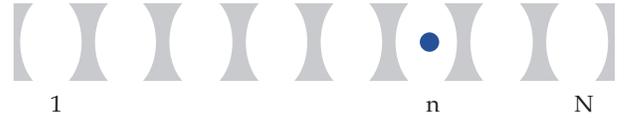}
  \caption{(Color online). Schematic configuration of our system.
   A two-level atom (filled blue circle) is in the $n$-th cavity
   of the CCA\@.}
  \label{fig:0}
\end{figure}

We consider a two-level atom located in the $n$-th cavity of a 1D
CCA with length $N$ as shown in Fig.~\ref{fig:0}, where the atom is
our quantum system and the CCA is its 1D electromagnetic environment.

Under the rotating wave approximation the Hamiltonian of the system
becomes
\begin{eqnarray}
  \mathcal{H}=\mathcal{H}_{0}+\mathcal{H}_{I}+\mathcal{H}_{A},
\end{eqnarray}
where  the Hamiltonian of the CCA $H_{0}$ is
\begin{equation}
  \mathcal{H}_{0} = \omega_c \sum_{j=1}^{N} a_{j}^{\dag} a_{j} - \eta
  \sum_{j=1}^{N} \left( a^{\dag}_{j} a_{j-1} + \mathrm{h.c.} \right)
\end{equation}
with $\omega_{c}$ the frequency of each single-mode cavity, $\eta$ the
hopping strength, $N$ the number of cavities in the CCA, and
$a^{\dag}_{j}$ ($a_{j}$) the photon creation (annihilation) operator
for the $j$-th single-mode cavity. Hereafter, we set $\eta=1$ and
$\omega_{c}=0$;  The Hamiltonian for the free two-level atom
$\mathcal{H}_{a}$ is
\begin{equation}
  \mathcal{H}_{a}=\omega_{a}|e\rangle\langle e|,
\end{equation}
where $\omega_a$ is the
energy of the excited state, $|e\rangle$ ($|g\rangle$) is the excited
(ground) state of the atom. The interaction between the atom
the cavity $\mathcal{H}_{I}$ is
\begin{equation}
  \mathcal{H}_{I}=g\left(a_{n}^{\dag}\sigma_{-} + a_{n}\sigma_{+}\right),
\end{equation}
where $g$ is the coupling the atom and the cavity, and
$\sigma^{-}$($\sigma^{+}$) is the atomic lowering (raising) operator.

Notice that the total photon number
$n_{p}=\sum_{j=1}^{N}a_{j}^{\dagger}a_{j}$ in $H_{0}$ is
conserved for the CCA\@. In the subspace of $n_{p}=1$, the eigenstate
and eigenvalue of the CCA is
\begin{eqnarray}
  |\varphi_{k}\rangle & = &  \sqrt{\frac{2}{N+1}} \sum^{N}_{j=1}
                            \sin(j\theta_{k})
                            a^{\dag}_{j}|vac\rangle=b_k^{\dag}|vac\rangle, \label{eq:1}\\
  \omega_{k} & = &  -2\cos \theta_{k},\label{e10}
\end{eqnarray}
where $k$ is integer with $1\leq k \leq N$,
$\theta_{k}=\frac{k \pi}{N+1}$, $b_k$ ($b_k^{\dag}$) the the photon creation 
(annihilation) operator for the $k$-th mode and $|vac\rangle$ is the vacuum state
of the cavity array.

Using Eqs.~\eqref{eq:1} and \eqref{e10}, we can rewrite the
Hamiltonian as
\begin{eqnarray}
  \mathcal{H} = \sum_{k=1}^{N}\omega_k
  b_{k}^{\dag}b_{k}+\sum_{k=1}^{N}g_k
  \left(b^{\dag}_{k}\sigma_{-}+\mathrm
  {h.c.}\right)+\omega_{a}|e\rangle\langle e|
\end{eqnarray}
with $g_k=\sqrt{\frac{2}{N+1}} \sin(n\theta_{k})g$ being the coupling
between the atom and the $k$-th mode and. Notice that this is a typical
spin-boson model to study quantum open system, and here the energy
spectrum $\omega_{k}$ of the bosonic environment and the coupling
strength $g_{k}$ are specified.

\section{Exponential decay of the naked atom state}

Now we study the spontaneous decay of the two-level atom in our
system. In other words, when the atom is initially prepared in the
excited state and the CCA stays in its vacuum state, what is dynamics
of the excitation probability of the atom? In this section, we will
focus on studying the conditions for the exponential decay of the
excited atom, and theoretically predict the decay rate.

Due to the excitation number is conserved in our system, the state at
any time $t$ can be expanded as
\begin{equation}
  |\Psi(t)\rangle= \big[\alpha(t)\sigma_{+} + \sum^{N}_{k=1} \beta_k(t)
  b^{\dag}_{j}\big]|vac\rangle|g\rangle,
\end{equation}
where $\alpha(t)$ and $\beta_k(t)$ represent the excitation probability
amplitude for the atom state and the $k$-th mode at time $t$ respectively.

Through the Schr\"{o}dinger equation
\begin{equation}
  i\frac{\mathrm{d}}{\mathrm{d} t}|\Psi(t)\rangle=\mathcal{H}
  |\Psi(t)\rangle,
\end{equation}
we get the following set of equations about the coefficients
\begin{eqnarray}
   i\dot{\alpha}(t)=-i\omega_a\alpha(t)-i\sum_{k=1}^{N}g_k \beta_k(t),\\
   i\dot{\beta_k}(t)=-i\omega_k\beta_k(t)-ig_k \alpha(t).
\end{eqnarray}
Making the transformation of parameters
\begin{eqnarray}
  \left\{
   \begin{aligned}
   \widetilde{\alpha}(t)=\alpha(t)e^{i\omega_a t}\\
   \widetilde{\beta_k}(t)=\beta_k(t)e^{i\omega_k t}\\\label{e11}
   \Delta_k=\omega_k-\omega_a
   \end{aligned}
   \right.
\end{eqnarray}
we simplify the equations of parameters as
\begin{eqnarray}
   \dot{\widetilde{\alpha}}(t)=-i\sum_{k=1}^{N}g_k
  \widetilde{\beta_k}(t)e^{-i\Delta_k t}\label{e1}\\
   \dot{\widetilde{\beta_k}}(t)=-ig_k\widetilde{\alpha}(t)
  e^{i\Delta_k t}   \label{e2}
\end{eqnarray}
Integrate Eq.~(\ref{e2}) then substitute it into Eq.~(\ref{e1}), we
obtain an equation only about $\widetilde{\alpha}(t)$
\begin{eqnarray}
   \dot{\widetilde{\alpha}}(t)=-\sum_{k=1}^{N}g_k^2\int_0^{t}dt'
  \widetilde{\alpha}(t')e^{-i\Delta_k(t-t')}. \label{e7}
\end{eqnarray}

The exponential decay can be achieved when the near resonant modes
are closely spaced in frequency and share almost the same coupling
strength with the atom, meanwhile the other modes far off resonance
are of little importance in the decay. To meet up with the
requirements, we can use a cavity array with length long enough and
put the atom in a selected location. Then the decay rate can be analytically
derived. In our model, it is
\begin{eqnarray}
\Gamma=\frac{4g^2}{\sqrt{4-\omega_0^2}} \label{e3}
\end{eqnarray}
with $\omega_0=2\cos{k_0}=\omega_a$ the energy of the resonant mode
(see Appendix for the detailed calculation).

\begin{figure}[!htbp]
  \includegraphics[width=8cm]{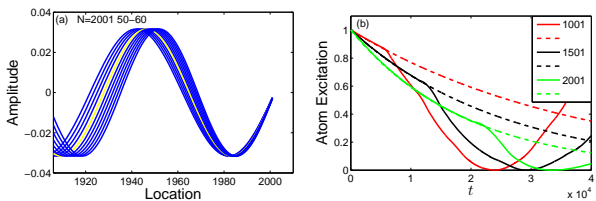}
  \caption{(Color online). (a) The $50$-th to $60$-th modes of CCA
  with $N=2001$. The yellow solid line stands
  for the mode ($55$-th) resonant with the atom while the other blue
  solid lines are for the modes nearby. (b) Time dependent atom state
  excitation for various lengths of CCA. The solid lines are
  the exact numerical results as for red $N=1001, n=992$, black $N=1501, n=1488$
  and green $N=2001, n=1984$. The corresponding dash lines are exponential
  decay lines with the decay rate determined by Eq.~(\ref{e3}). Here, $g=0.0015$
  and the atom is resonant with the $55$-th mode in all situations. }
  \label{fig:1}
\end{figure}
We now give some numerical results to support the theoretical analysis above.
Figure~\ref{fig:1}(a) shows the modes condition around the resonant frequency.
With the atom set in the $1984$-th cavity (the antinode of the resonant mode),
the above mentioned requirements are satisfied. Furthermore, in a short time
the behavior of the atom state is a truly well-defined exponential decay and
with its decay rate well predicated by Eq.~(\ref{e3}) (see Figure~\ref{fig:1}(b)).

But when time passes a specific point, the evolution dramatically
deviates from the exponential line, and the change can be postponed
by lengthening the cavity array. Next, we will
explain the appearance of the turning point, and show that after the point a
new physical process starts.

\section{Deviating from the exponential decay}

\begin{figure}[!htbp]
  \includegraphics[width=8cm]{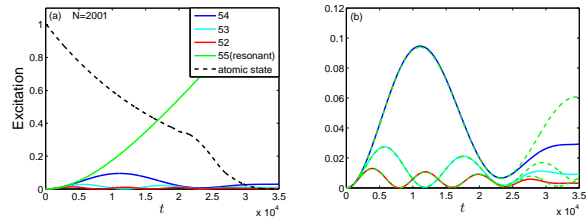}
  \caption{(Color online). (a) Black dash line is the atom excitation
    and solid lines stand for various modes. (b) Only the
    near-resonance modes, for the solid lines each color represents
    the same mode as in (a), the yellow dashed line stands for the
    results calculate from Eq.~(\ref{e4}). To simplify the figure, we
    only show some blue tuned near-resonance modes while not including
    the red tuned ones, their behavior actually are very much alike. }
  \label{fig:2}
\end{figure}

The key point for understanding the deviation is to clarify the
behavior of the modes around the resonant frequency. We pick out the
$N=2001$ condition for example, Fig~\ref{fig:2}(a) reveals the
evolution of the modes alongside with the atom state. Around the
turning point, except the resonant mode other near-resonance modes
almost all together oscillate to near zero. Fig~\ref{fig:2}(b) gives a
more clear view about this pattern. This is the main reason to cause
the deviation since the resonant mode becomes predominate over other
near resonant modes at the turning point, which violates the
requirements for the exponential decay of a naked atom.

Now we turn to give analytical analysis about the appearance of this pattern.
As the atom state exponentially decay with time, its excitation probability
amplitude is
\begin{equation}
\widetilde{\alpha}(t)=e^{-\Gamma t/2},
\end{equation}

Then the $k$-th mode's excitation probability amplitude can be
determined through Eq.~(\ref{e2})
\begin{equation}
\begin{aligned}
\widetilde{\beta_k}(t)&=-i g_k\int^{t}_{0}dt' e^{(i\Delta_k-\Gamma/2)t'}\\
&=\frac{-ig_k}{i\omega_k-\Gamma/2}[e^{(i\Delta_k-\Gamma/2)t}-1],
\end{aligned}
\end{equation}
while its probability is
\begin{equation}
M_k(t)=|\widetilde{\beta_k}(t)|^2=A_k(e^{-\Gamma t}-2e^{-\Gamma t/2}\cos\Delta_k t+1),\label{e4}
\end{equation}
with $A_k=\frac{g_k^2}{\Gamma^2/4+\omega_k^2}$.

Eq.~(\ref{e4}) can perfectly describe the evolution of the
modes during the time when the atomic state obeys the exponential decay
as shown in Fig~\ref{fig:2}(b). For $\Delta_k$, it satisfies the
following equation $\Delta_k\approx(k-k_0)\Delta_1$ (see Appendix for
proof), here we define $\Delta_1=\omega_{k_0+1}-\omega_{k_0}$. To determine
the minimum of $M_k$, we calculate its derivative
\begin{equation}
  \dot{M_k(t)}=A_ke^{-\Gamma t/2}(-\Gamma e^{-\Gamma t/2}+\Gamma
  \cos\Delta_k t+2\Delta_k\sin\Delta_k). \label{e8}
\end{equation}
For $\Delta_k\gg \Gamma$, only the third term in the right-hand side
of Eq.~(\ref{e8}) is predominant. Then minimums appear near
\begin{equation}
t_k=\frac{2\pi l}{\Delta_k}\approx \frac{2\pi l}{|k-k_0|\Delta_1}\label{e9}
\end{equation}
as $l$ is positive integer.

With Eq.~(\ref{e9}) we can see when the excitation of the nearest mode (mode $k_0\pm 1$)
reach the first minimum, the other near-resonance modes also reach its
minimum (though they have already oscillated several times). Then the time of
the turning point $t_c$ is totally dependent on the energy difference between
the resonant and nearest near-resonance mode
\begin{equation}
t_c=\frac{2\pi}{\Delta_1}
\end{equation}
And this is well consistent with the numerical results. When lengthening
the cavity array, $\Delta_1$ decreases and leads to the postpone
the turning point as we mentioned before. For the extreme case,
a 1D CCA with infinite length $\Delta_1\rightarrow 0$ leads to $t_c\rightarrow \infty$.
So this deviation phenomenon is actually due to the finite size of
the environment.

\section{Decay of the dressed atom}

Around the turning point, the effect of the resonant mode will be
predominant since the other near-resonance modes are almost empty
without any photon. We conclude that the atom and the resonant mode
will be strongly coupled to form a subsystem, the dressed
atom~\cite{3}. So for the next stage of time, it is the decay of the
dressed atom, a physical process totally different with before.

We can introduce a general master equation to describe the evolution of
the dressed atom~\cite{3}
\begin{equation}
  \frac{{\rm d}}{{\rm d}t}\rho = -i[H_d, \rho] - \frac{\Gamma}{2}
  (\sigma^{+}\sigma^{-}\rho_d+\rho\sigma^{+}\sigma^{-}) +
  \Gamma\sigma^{-}\rho\sigma^{+}
\end{equation}
with
\begin{equation}
  H_d = \frac{1}{2} \omega_0\sigma_z + \omega_0 b^{\dag}_r b_r + g_r
  (b^{\dag}_r\sigma^{-}+h.c.) \label{e5}
\end{equation}
the Hamiltonian of the dressed atom,
$g_r$ $(=\sqrt{\frac{2}{N+1}}g)$ the coupling strength between the atom and
the resonant mode, $\rho$ the density matrix of the dressed atom and
$\Gamma$ the decay rate of the subsystem (same as the the spontaneous
emission rate of the atom).

Under the secular approximation, we can project the master equation
over the dressed states
\begin{equation}
  \frac{{\rm d}}{{\rm d}t}\rho_{ii}=-\frac{\Gamma}{2}\rho_{ii}\label{b2}
\end{equation}
\begin{equation}
\frac{{\rm d}}{{\rm d}t}\rho_{12} =
(-2ig_r-\frac{\Gamma}{2})\rho_{12}\label{b3}
\end{equation}
with $i=1,2$, $\rho_{ij}=\langle i|\rho|j\rangle$,
$|1\rangle=\frac{1}{\sqrt{2}}
(|vac\rangle|e\rangle+|1_{k_0}\rangle|g\rangle)$
and
$|2\rangle=\frac{1}{\sqrt{2}}
(|vac\rangle|e\rangle-|1_{k_0}\rangle|g\rangle)$ the dressed states.

\begin{figure}[!htbp]
  \includegraphics[width=8cm]{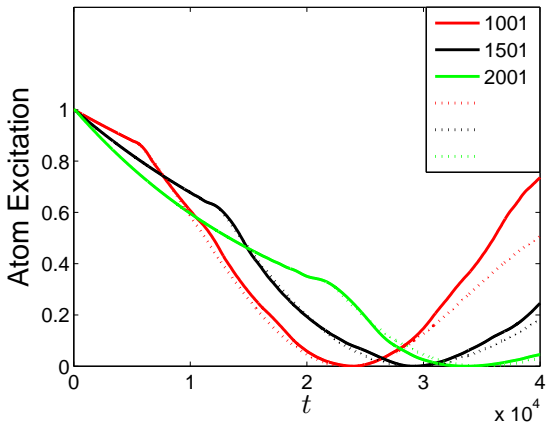}
  \caption{(Color online). Atom excitation as a function
  of time. Solid lines represent exact numerical results
  with different color stands for different cavity array length.
  The dotted lines are obtained through Eq.~(\ref{e6}). All the
  parameters are the same as before.  }
  \label{fig:3}
\end{figure}

For the atom excitation
\begin{equation}
\rho_{ee}(t)=\frac{1}{2}[\rho_{11}(t)+\rho_{22}(t)+\rho_{12}(t)+\rho_{21}(t)]\label{b1}
\end{equation}
Substitute the results obtained through Eq.~(\ref{b2}) and Eq.~(\ref{b3}) into
Eq.~(\ref{b1}), the evolution of the atom state is determined as
\begin{equation}
\begin{aligned}
\rho_{ee}(t)&=\frac{1}{2}e^{-\frac{\Gamma}{2}(t-t_c)}[\rho_{11}(t_c)+\rho_{22}(t_c)\\
&+\rho_{12}(t_c)e^{-2ig_r(t-t_c)}+\rho_{21}(t_c)e^{2ig_r(t-t_c)}]\label{e6}
\end{aligned}
\end{equation}
with $\rho_{ee}(t)=\langle e|\langle vac |\rho(t)|vac\rangle|e\rangle$.

Fig~\ref{fig:3} shows that after turning point, the dressed atom decay
theory can describe the system evolution quite accurately. But after
the time point when the atom state is empty, the dressed atom system
no longer exists, the evolution becomes quite chaotic and unpredictable.

\vskip 0.5cm
In conclusion, we study the atom decay within the 1D CCA frame. By
selecting appropriate parameters, we achieve the well-known
exponential decay of the atom state, and theoretically give the decay
rate. Meanwhile the exponential decay only last a limited time in this
model, the evolution deviates after a time point we name $t_c$. This
is due to the near-resonance modes almost become empty together around
$t_c$, and we theoretically prove the existence of $t_c$ in our model.
After $t_c$, the resonant mode is strongly coupled with the atom to
form the dressed atom subsystem, and the dressed atom decay theory can
describe the evolution quite accurately. Finally when the atom evolves
into its ground state, the dressed atom subsystem no longer exists,
and the evolution after is quite chaotic and beyond our understanding
now. We hope our work can enlighten the study of the transition
between the atom state exponential decay and dressed atom decay, and
help to have a better understanding about the dynamics of quantum open
system with a finite size environment.

\begin{acknowledgments}
  This work is supported by NSF of China (Grants No. 11475254 and
  Grant No. 11175247) and NKBRSF of China (Grant Nos. 2012CB922104 and
  2014CB921202).
\end{acknowledgments}

\section*{APPENDIX: How many modes are enough?}

Throughout the paper, we emphases that only modes around the resonant
frequency are important in the system evolution while the far
off-resonance modes can be ignored (this is also the key point in the
Weisskopf-Wigner approximation). Then the question appears immediately,
how to define near-resonance and far off-resonance or in another way
say how many modes is actually enough? We give some numerical results
to show intuitively that the number is quite small.

\begin{figure}[!htbp]
  \includegraphics[width=8cm]{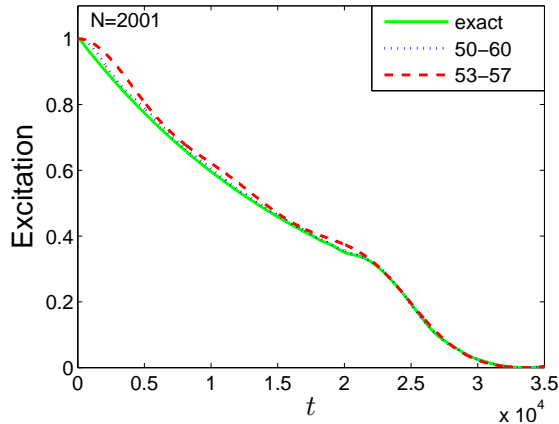}
  \caption{(Color online). Time dependent atom excitation. The green
    solid line is the exact results including all the modes, the
    dotted blue line including the $50$-th to $60$-th modes and the red
    dashed line including the $53$-th to $57$-th modes. Here the other
    parameters are the same as before. }
  \label{fig:4}
\end{figure}

For the cavity array with $N=2001$, Fig~\ref{fig:4} reveals that the
system can be described quite accurately using only $5$ modes and
perfectly with $11$ modes. Compared with the total $2001$ modes in the
system, this is a very small number.

\section*{APPENDIX: CALCULATION OF THE DECAY RATE}

As mentioned above, for the modes around the resonant frequency the
coupling strength $g_k\approx g_r=\sqrt{\frac{2}{N+1}}g$, then we can
rewrite Eq.~(\ref{e7}) as
\begin{eqnarray}
   \dot{\widetilde{\alpha}}(t) = -\frac{2g^2}{N+1} \sum_{k=1}^{N}
  \int_0^{t}\mathrm{d}t' \widetilde{\alpha}(t')e^{-i\Delta_k(t-t')}.
\end{eqnarray}
After replacing the summation over $k$ with an integration as ${\rm
  d}k=\frac{\pi}{N+1}$, we get
\begin{eqnarray}
   \dot{\widetilde{\alpha}}(t) = -\frac{2g^2}{\pi}
  \int^{\pi}_{0}\mathrm{d}k\int_0^{t}\mathrm{d}t'
  \widetilde{\alpha}(t')e^{-i\Delta_k(t-t')}.
\end{eqnarray}
By replacing the integral variable $k$ with frequency $\omega$ since
$\omega=2\cos{k}$, we have
\begin{eqnarray}
   \dot{\widetilde{\alpha}}(t)=-\frac{2g^2}{\pi}\int_0^{t}\mathrm{d}t'\int^{2}_{-2}\mathrm{d}\omega
  \frac{1}{\sqrt{4-\omega^2}}\widetilde{\alpha}(t') e^{i(\omega_0-\omega)(t-t')}.
\end{eqnarray}
Adopting the Weisskopf-Wigner approximation~\cite{1,x1}, we get
\begin{eqnarray}
  \dot{\widetilde{\alpha}}(t) = -\widetilde{\alpha}(t)
  \int_0^{\infty}\mathrm{d}\tau\int^{2}_{-2} \mathrm{d}\omega\frac{2g^2}{\pi\sqrt{4-\omega^2}}
  e^{i(\omega_0-\omega)\tau}
\end{eqnarray}
with $\tau=t-t'$. Since
\begin{eqnarray}
\int^{\infty}_{0}\mathrm{d}\tau e^{-i(\omega-\omega_0)\tau}=\pi\delta(\omega-\omega_0)-iP(\frac{1}{\omega-\omega_0})
\end{eqnarray}
where $P$ represents the Cauchy principal part which leads to the
well-known lamb shift. Since the lamb shift has nothing to do with
decay, we neglect it here and obtain
\begin{eqnarray}
   \dot{\widetilde{\alpha}}(t)= e^{-\frac{\Gamma }{2}t}
\end{eqnarray}
with $\Gamma$ the atom spontaneous emission rate
\begin{eqnarray}
\Gamma=\frac{4g^2}{\sqrt{4-\omega_0^2}}
\end{eqnarray}

\section*{APPENDIX: Proof of $\Delta_k\approx(k-k_0)\Delta_1$}

From Eq.~(\ref{e9}) and Eq.~(\ref{e10}), we have
\begin{eqnarray}
\Delta_k=2(\cos{\frac{k\pi}{N+1}}-\cos{\frac{k_0\pi}{N+1}}).
\end{eqnarray}
Using the sum and difference formulas of the trigonometric functions,
we rewrite it as
\begin{eqnarray}
\Delta_k=-4\sin{\frac{m\pi}{2(N+1)}\sin{\frac{2k_0+m}{2(N+1)}}}, \label{e12}
\end{eqnarray}
where  $m=k-k_0$.
Since we only consider the modes around the resonant frequency,
$m\ll N$ and $m\ll k_0$. Then Eq.~(\ref{e12}) can be simplified as
\begin{eqnarray}
\Delta_k\approx-4\frac{m\pi}{2(N+1)}\sin{\frac{2k_0+m}{2(N+1)}}\approx (k-k_0)\Delta_1.
\end{eqnarray}

\end{document}